\documentclass[twocolumn,pra,aps,superscriptaddress,showpacs, flushbottom]{revtex4-1}
\usepackage{xspace,amsmath,amsfonts,amsthm,amssymb,amsbsy,graphicx,color}

\newcommand{\ms}[1]{\mbox{\scriptsize #1}}

\begin{document}

\newtheorem{theo}{Theorem} \newtheorem{lemma}{Lemma}

\title{Quantum-enhanced accelerometry with a non-linear electromechanical circuit}

\author{Kurt Jacobs} 
\affiliation{U.S. Army Research Laboratory, Computational and Information Sciences Directorate, Adelphi, Maryland 20783, USA} 
\affiliation{Department of Physics, University of Massachusetts at Boston, Boston, MA 02125, USA}
\affiliation{Hearne Institute for Theoretical Physics, Louisiana State University, Baton Rouge, LA 70803, USA} 
\author{Radhakrishnan Balu}
\affiliation{U.S. Army Research Laboratory, Computational and Information Sciences Directorate, Adelphi, Maryland 20783, USA} 
\author{John D. Teufel}
\affiliation{National Institute of Standards and Technology, Boulder, CO 80305, USA}

\begin{abstract}   
It is known that placing a mechanical oscillator in a superposition of coherent states allows, in theory, a measurement of a linear force whose sensitivity increases with the amplitude of the mechanical oscillations, a uniquely quantum effect. Further, entangled versions of these states across a network of $n$ mechanical oscillators enables a measurement whose sensitivity increases linearly with $n$, thus improving over the classical scaling by $\sqrt{n}$. One of the key challenges in exploiting this effect is processing the signal so that it can be readily measured; linear processing is insufficient. Here we show that a Kerr oscillator will not only create the necessary states, but also perform the required processing, transforming the quantum phase imprinted by the force signal into a shift in amplitude measurable with homodyne detection. This allows us to design a relatively simple quantum electro-mechanical circuit that can demonstrate the core quantum effect at the heart of this scheme, amplitude-dependent force sensitivity. We derive analytic expressions for the performance of the circuit, including thermal mechanical noise and photon loss. We discuss the experimental challenges in implementing the scheme with near-term technology. 
\end{abstract} 

\pacs{03.67.-a, 85.85.+j , 42.50.Dv, 85.25.Cp} 

\maketitle 

The uniquely quantum-mechanical properties of mesoscopic systems hold the potential to improve sensors far beyond what is possible with classical devices. This potential is one of the driving forces in the development of controllable quantum systems
~\cite{Gannaway78, Caves81, Bollinger96,  Huelga97, Giovannetti01b, Treps02, Giovannetti04, Kok04, Giovannetti06b, Higgins07, Dowling08, Boixo08b, Berry09, Wasilewski10, Schleier10, Tsang10, Vuletic12, LIGO13, Woolley13, Bohnet14, Komar14, Taylor16,  Jacobs16}. Here we show how a mesoscopic device, consisting of two superconducting nonlinear oscillators coupled to a nano-mechanical resonator, can be used to harness the ability of superpositions of coherent states to perform accelerometry in a uniquely quantum-enabled manner.  Quantum electromechanical circuits with the required components have already been demonstrated in experiments~\cite{Teufel11, Teufel11b, Palomaki13, Palomaki13b, Lecocq15}. 

The measurement of a linear force using a classical oscillator proceeds by detecting the shift in momentum, $\Delta p$, induced by the force, $F$, which is achieved by measuring the resulting shift in position. If the force is resonant with the oscillator and acts for a time $\Delta t$, then the momentum shift is $\Delta p = F \Delta t$, and when translated to a shift in position via a quarter period of the oscillation is $\Delta x = F \Delta t/(m\omega)$. Given the uncertainty in a coherent state of a quantum oscillator, a standard classical (coherent-state) method for measuring position has a single-shot fractional error of $\varepsilon_{\ms{c}} = \sigma_x/\Delta x = \sqrt{\hbar m\omega/2}/(F \Delta t)$, in which $m$ and $\omega$ are respectively the mass and frequency of the mechanical oscillator. This error assumes the limit in which the position measurement itself is has an error much less than $\sigma_x$ (implying, for example, that an interferometer used to measure the position employs a sufficiently large number of photons). 

It turns out that a remarkable quantum effect, first noted in~\cite{Munro02}, makes it possible to increase the sensitivity of an oscillator to an applied linear force by placing the oscillator in a superposition of two coherent states. Such states, in which a mechanical oscillator is in ``two places at the same time'', are often referred to as ``cat'' states, and have been generated in electrical oscillators using the Kerr nonlinearity~\cite{Kirchmair13, Vlastakis13}. The origin of the effect is the global phase imparted to a coherent state by the action of a force, a phase that becomes observable only when one has superpositions of coherent states. The action of a linear force $F$ acting for a time $\Delta t$ is described by $V(F) = e^{-i F x \Delta t/\hbar}$ where $x$ is the position operator. Using the scaled position defined by $\tilde{x} = a + a^\dagger$, we can write this as $V = e^{-i \delta \tilde{x}}$, in which 
    $\delta  =   F \Delta t/\sqrt{2m\omega\hbar}$    
is a scaled (dimensionless) version of the momentum kick from the force. The action of $V$ on a coherent state $|\alpha_0\rangle$ is $V|\alpha_0\rangle = e^{-i\delta\alpha} |\alpha_0+i\delta\rangle$, in which we have defined $\alpha \equiv \mbox{Re}[\alpha_0]$. The action of a force on the cat state  
\begin{align} 
   |\alpha_0\rangle_{\ms{c}} \equiv \left[ |\alpha_0 \rangle + i  |\! -\!\alpha_0\rangle \right]  /\sqrt{2}
   \label{x2}
\end{align}
is therefore   
\begin{align}
  V(\delta) |\alpha_0\rangle_{\ms{c}} = \left[ |\alpha_0 + i \delta\rangle + i e^{i2 \delta\alpha} |\! -\!\alpha_0 + i\delta\rangle \right]  /\sqrt{2} .  \label{x1}
\end{align} 
We see that the phase shift generated in the superposition by the force is equal to $2 \delta\alpha$, and hence proportional to the real part of the coherent amplitude $\alpha_0$. 

Since the signal appears as a phase shift between two nearly orthogonal states it is essentially a rotation in a two-dimensional state-space. Because of this each single-shot measurement of this state-space gives just 1 bit of information regarding the signal. As discussed in more detail below, the resulting fractional error for a single shot is $\varepsilon_{\ms{q}} = \sqrt{m\omega\hbar/2} /(2\alpha F \Delta t) = \varepsilon_{\ms{c}}/(2\alpha)$. It must be noted that while this quantum error can be made much smaller than the classical value by using $\alpha \gg 1$, it is limited in that it cannot be smaller than approximately unity (see below). Thus the quantum method beats the classical method per shot when the force to be measured is sufficiently small that the classical single-shot error is much larger than unity, meaning that the momentum shift $\Delta P = F \Delta t \ll \sqrt{m\omega\hbar/2}$. 

To date, one other concrete scheme for exploiting a cat state for force measurement has been devised, and this uses an ion trap~\cite{Toscano06, Dalvit06}. The primary advance represented by our work here is the nonlinear processing we introduce and the resulting simplification. We note that squeezed states also provide a way to realize quantum-enhanced sensitivity at the Heisenberg limit (meaning a sensitivity that scales as the square root of the mechanical energy~\cite{Luis04, Hall12}). Nevertheless, cat states have a number of potential advantages over squeezed states: cat states can be prepared in a time that is independent of amplitude, the enhancement from squeezed states is limited by the degree of squeezing, and the cat-state method can be generalized to provide an additional quantum enhancement of $\sqrt{N}$ by using entangled cat states of $N$ oscillators. It is also important to note that the (quantum enhanced) scaling of sensitivity with phonon number that we consider here is quite distinct from that of the precision of an interferometer (in that case with photon number). The latter scaling is often discussed in the context of force measurement (e.g. LIGO~\cite{McKenzie02, LIGO13, Miller15}) because interferometers are used in this context to make position measurements on mechanical oscillators. The pulsed measurement method we consider here does not use either a position measurement or an interferometer. As such the resources involve the mechanical amplitude instead of that of the auxiliary cavity modes of interferometric schemes. 

We now show that a Kerr nonlinearity~\cite{Mancini97, Bose97, Jacobs14} will translate the force signal --- the phase of the superposition ---  into a shift of the average value of $X = (a + a^\dagger)/2$. We will then discuss how this shift is estimated from the measurement results and the resulting single-shot error, before turning to our electromechanical circuit. We apply the Kerr evolution $U_{\lambda t} = \exp [-i \lambda t (a^\dagger a)^2]$ for $\lambda t = \pi/2$ to a coherent state $|\alpha_0\rangle$ to generate the cat state in Eq.(\ref{x2}), followed by the force $V(\delta)$, resulting in the state that carries the force signal (Eq.(\ref{x1})). We then apply the inverse Kerr evolution $U_{\pi/2}^\dagger$ (equivalently $U_{3\pi/2}$) and calculate the average value of $\langle X \rangle$:  
\begin{align} 
  \langle X \rangle & =  \mbox{Re}[\langle \alpha_0 | U^\dagger V^\dagger U a U^\dagger  V U  |\alpha_0\rangle]  \nonumber \\ 
   &   =  e^{-2\delta^2}  \left\{ \alpha \cos(4 \alpha \delta) - \delta \left[\sin(4 \alpha \delta) - e^{-2\alpha^2} \right]  \right\}  .    \label{S1}
\end{align}
Here we have defined $U \equiv U_{\pi/2}$ and once again $\alpha \equiv \mbox{Re}[\alpha_0]$. We perform the above calculation by first applying $UV$ to the state $|\alpha_0\rangle$, and then using $UaU^\dagger = i a \exp(i \pi a^\dagger a)$ and the fact that $\exp(i k a^\dagger a) |\alpha-)\rangle = |\alpha_0 e^{-ik}\rangle$. 

The expression for $\langle X \rangle$ does not have a linear dependence on the force, but we can obtain one by applying an ``offset'' force that shifts $\delta$ by $\pi/(8\alpha)$. Restricting the size of the measured force so that $\alpha \delta \ll 1$, and allowing $\alpha$ to be large enough to neglect $e^{-2\alpha^2}$, we obtain   
\begin{align} 
  \langle X \rangle & =  \alpha (4\alpha \delta) -  \delta   +  \mathcal{O}[(4\alpha\delta)^2]   ,  \;\;\;\;  4\alpha \delta \ll 1 . 
\end{align} 
Here we have separated out the two contributions of the amplitude of the initial coherent state, $\alpha$. The factor of $\alpha$ inside the parentheses is the (Heisenberg limited) quantum enhancement to the sensitivity of the accelerometer. The factor outside the parentheses does not contribute to the sensitivity as we now explain. The probability distribution for $X$ is peaked (approximately) at the two values $\pm\alpha$. The nonzero value of $\langle X \rangle$ is due to a difference between the probability of $ X = \alpha$ vs. that of $X = -\alpha$, and it is the probability imbalance that caries the signal. As such, one of the factors of $\alpha$ in $\langle X \rangle$ is due merely to the location of the peaks (and not the probability imbalance). The estimation process is just that of determining the probability of an unfair coin: we repeat the measurement process $M$ times, and the signal is 
\begin{align} 
  S = \frac{m}{M} - \frac{1}{2} = 2 \alpha  \left[ 1  -  \frac{1}{(2\alpha)^2} \right]  \delta, 
\end{align} 
in which $m$ is the number of times we obtain the result $X > 0$.  The error in the measurement of $S$ is approximately $\sigma_S = 1/\sqrt{4M}$ and the single-shot error is $\sim 1/2$. 

Two further properties of the above scheme are worth noting. The first is the way in which the force appears in the phase shift, $4\alpha \delta$. Since this phase shift is proportional to $\mbox{Re}[\alpha_0]$, and complex number $\alpha_0$ rotates at the mechanical frequency, the detection scheme filters the force signal: a constant force produces no signal if it acts for an integral number of periods, whereas the shift produced by a force at the mechanical frequency simply grows with time. The scheme thus filters the force in a band around the mechanical frequency, where the bandwidth is inversely proportional to the time the state spends in the mechanics. This filter is also nonlinear. The second property is that the filter is phase-selective: because the phase shift is proportional to the real part of $\alpha_0$, it primarily detects the quadrature that is in phase with the resonator. To measure both quadratures one could either use two devices, or shift the phase of the mechanical oscillator periodically. 

\begin{figure}[t] 
\leavevmode\includegraphics[width=1\hsize]{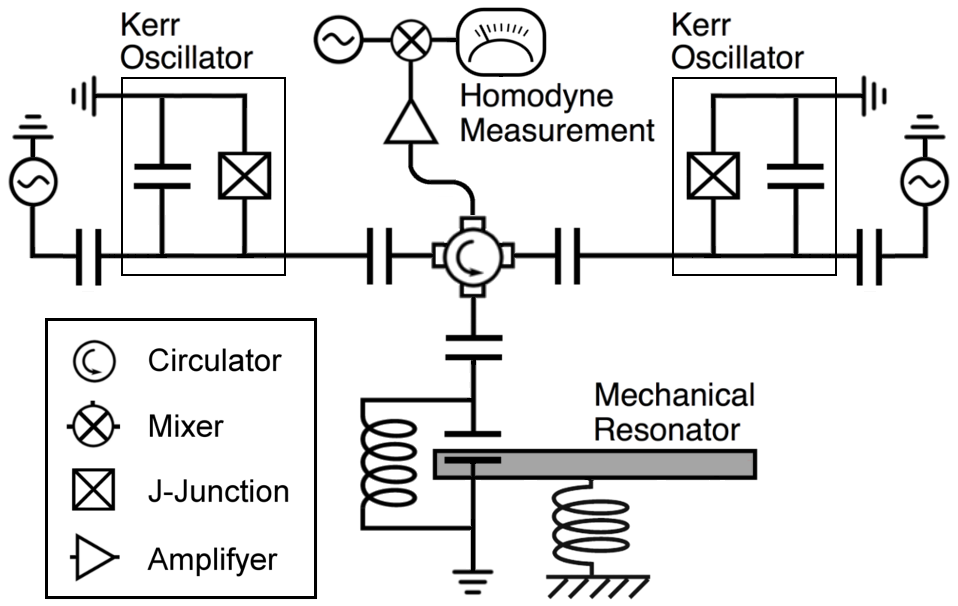} 
\caption{An electromechanical circuit that implements a quantum-enhanced measurement of a force using a superposition of two coherent states. The functioning of the circuit is described in the text.} 
\label{fig1} 
\end{figure}

The circuit we suggest for implementing the above scheme is depicted in Fig.\ref{fig1}. The nonlinear electrical oscillator on the left hand side in Fig.1 (constructed using a Josephson junction~\cite{Vion02, Koch07, Jacobs14}) creates a cat and transfers this to the mechanical oscillator. This transfer is accomplished via a transmission line that is coupled to the mechanics via an ``auxiliary'' linear electrical oscillator. The cat state picks up the force signal and is then transferred to a second nonlinear electrical oscillator (on the right in Fig.1) via the auxiliary oscillator and transmission line. Finally the second Kerr oscillator is measured via homodyne detection. 
We note that a single Kerr oscillator would suffice if we eliminated the transmission line and used a direct interaction between this oscillator and the auxiliary. But in that case the linear oscillator would inherit some of the nonlinearity, a situation that is problematic for the transfer.  

%


We now analyze the functioning of the circuit in Fig.~\ref{fig1}, with the primary purpose of examining the effects of thermal noise in the mechanical oscillator and photon loss (damping) in the superconducting components. It turns out that analytic expressions can be derived when the number of photons lost from the Kerr oscillators remains small. Preparing the initial coherent state in the Kerr oscillator is fairly straightforward, so we consider the evolution as having three parts: I) the nonlinear Kerr evolution that creates the cat state $|\alpha_0\rangle_{\ms{c}}$, II) the linear evolution that transfers this state to the mechanics and back into the Kerr oscillator, and III) the final Kerr evolution prior to the homodyne measurement. 

For the purposes of analysis we can treat the two nonlinear oscillators as a single oscillator that transfers its state to the mechanics and subsequently receives the state back. The transfer, implemented via a transmission line and an intermediate linear oscillator coupled to the mechanics, is a linear transformation. In order to obtain analytic results we model this linear transfer (part II of the evolution) using the linear ``swapping'' interaction $H_{\ms{I}} = \hbar g (a b^\dagger + b a^\dagger)$, in which $a$ and $b$ are, respectively, the annihilation operators for the Kerr oscillator and mechanical mode. Here $g$ sets the time taken for the transfer, and incorporates the coupling rates of the Kerr and auxiliary oscillators to the transmission line and the coupling of the mechanics to the auxiliary. We note that the precise time-dependence of the transfer process implemented by the circuit (see, e.g.~\cite{Cirac97}) will differ somewhat from the cosine evolution of $H_{\ms{I}}$, and we discuss the expected effects of this difference briefly below. The transfer due to $H_{\ms{I}}$, including damping and thermal noise for the mechanical mode with rate $\gamma$ at temperature $T$, internal damping of the Kerr and auxiliary oscillators at the combined rate $\kappa$, and the force applied to the mechanics is described by the linear quantum Langevin equations of input-output theory~\cite{Gardiner85, Jacobs14}. Solving these equations one finds that the frequency of the swapping process is $\nu = \sqrt{g^2 - (\kappa + \gamma)^2/16}$, and the time to complete a single swap into and back out of the mechanical mode is $T = \pi/\nu$. At the completion of the swapping process, in the frame rotating at the frequency of the Kerr oscillator, the mode operator $a$ is given by 
\begin{align} 
 a(T) =  e^{-\Gamma T} a(0) e^{-i \omega T} + D_{\ms{in}} + i \delta', 
\end{align}
in which $a(0)$ is the mode operator at the start of the transfer, $\omega$ is the frequency of the mechanical mode, $\Gamma = (\kappa + \gamma)/4$, $D_{\ms{in}}$ is a time-integral of the vacuum field inputs to the oscillators, and $\delta'$ is the dimensionless version of the total momentum kick (the expression for which is given below in Eq.(\ref{del}). The fact that the coefficient of $a(0)$ is less than unity describes the loss of photons into the vacuum environment. Additional loss of amplitude due to the transmission line by replacing the damping rate of the Kerr oscillator by an increased effective rate $\kappa'$. Calculating the signal is simplified greatly by defining a unitary operator $L$ that generates the evolution $a(0) \rightarrow a(T)$ in terms of an auxiliary mode that absorbs the lost photons. This operator is 
\begin{align} 
  L = \exp [ \xi c a^\dagger  - \xi^* ac^\dagger] \exp[ -i(\delta'/z)a^\dagger - i(\delta'/z)^* a ]   
\end{align} 
in which $c$ is the annihilation operator for the auxiliary mode and $\xi = e^{-\Gamma T}$. 

To obtain analytical results for the nonlinear part of the evolution we employ the method of ``linear trajectories''. In this method the mixed-state evolution is explicitly broken into a sum over various pure-state evolutions (``trajectories'') corresponding to the emission (loss) of different numbers of photons~\cite{Jacobs14}. For each trajectory the normalization of the evolved state caries information about the probability with which that trajectory occurs. Since the evolution of interest in parts I and III involves only the Kerr oscillator the sole non-unitary component is photon loss. 

The linear trajectory in which no photons are lost by the Kerr oscillator is obtained merely by replacing $U_{\lambda t}$ by the operator $W_{\lambda t} = U_{\lambda t} e^{-\kappa t a^\dagger a/2}$ and normalizing the final state. Here we are interested in the error in the measurement caused by the loss of one or more photons, and $W_{\lambda t}$ provides the probability that such a loss occurs. Further quantitative insight into the effect of photon loss is obtained from the evolution operator for the trajectory in which a single photon is emitted at time $t'$, namely $W_1(t)  =  W(t - t') a W(t')$. This reveals that each emission kicks the phase of the cat by a random amount distributed symmetrically over the interval $[0,\pi]$. 

Now that we have captured the evolution in terms of relatively simple operators $W$ and $L$, we first calculate the signal in the absence of photon emissions. The initial state of the Kerr oscillator is $|\alpha_0\rangle$ and that of the auxiliary is the vacuum. That of the resonator can be omitted since it has no effect on the process. Writing the initial joint state as $|\alpha_0\rangle |0\rangle$, the expression for $\langle X \rangle$ is 
\begin{align} 
  \langle X \rangle_0 & = \mathcal{N}^{-1}\mbox{Re} [ \langle 0 | \langle \alpha_0 | W^\dagger L^\dagger W \, a\, W^\dagger  L W  |\alpha_0 \rangle |0\rangle ]  
\end{align} 
in which $\mathcal{N}$ is the required normalization. The apparent simplicity of the above expression is somewhat deceptive: it is considerably more challenging to evaluate than Eq.(\ref{S1}). We proceed by applying $LW$ to $|\alpha \rangle |0\rangle$ which produces a superposition of unnormalized coherent states. We then use $W \, a\, W^\dagger = i e^{-\kappa t a^\dagger a} a e^{-i\pi a^\dagger a}$ and apply this expression to the superposition, employing the fact that $\exp(\zeta a^\dagger a) |\alpha \rangle = e^{|\alpha|^2 (\eta - 1)} |\alpha \eta \rangle$ with $\eta = e^{\zeta}$. The normalization is given by $\mathcal{N} = \langle \psi | \psi \rangle$ with $| \psi \rangle = W^\dagger  L W |\alpha \rangle |0\rangle$. In the regime of low loss, $\gamma\tau \lesssim \kappa\tau  \ll  1$, we obtain   
\begin{align*} 
  \langle X \rangle_0 & = \eta e^{-\eta |\delta'|^2} [ -  \xi \eta \alpha \cos (4 \eta^2 \alpha \delta') - \delta' \sin(4 \eta^2 \alpha \delta')] , 
\end{align*} 
in which $\eta \approx 1 - (\gamma \tau)/2$, $\xi = e^{-\Gamma T}$, and we have dropped the part of $S_0$ that is suppressed by the factor $e^{-2\alpha^2}$ ($\alpha \gg 1$). Applying the phase offset of $90^\circ$ and taking $\alpha\delta' \ll 1$ gives   
\begin{align} 
   \langle X \rangle_0 & = - \xi \eta^2  \alpha (4 \eta^2 \alpha \delta') - \eta \delta' ,  \label{xx2}
\end{align}
in which the dimensionless momentum kick is 
\begin{align} 
  \delta' =  \mbox{Re} \left[   \int_0^{\pi/\nu} \sin(\nu s) e^{i\omega s} f(s) +  \tilde{n}(s) ds   \right] .    \label{del} 
\end{align} 
Here $f(t) = \sqrt{2} F(t)/\sqrt{\hbar\omega m}$ is the scaled force, $\tilde{n}(t)$ is the thermal noise, and $\nu^2 = g^2 - (\kappa - \gamma)^2/16$. The correlation function of the thermal noise is $\langle \tilde{n}(t) \tilde{n}(t') \rangle =  [2 n(\omega) + 1]\delta(t'-t)$, with $n(\omega) = [e^{kT/(\hbar\omega)} - 1]^{-1}$ in which $T$ is the ambient temperature and $k$ is Boltzmann's constant. As noted above, the time envelope of the transfer will differ in practice from the simple cosine evolution that we have employed here. We can see that the effect of this will be primarily to modify the sine envelope in the integral for $\delta'$, and to slightly modify the relative contributions of $\kappa$ and $\gamma$ to the decoherence. 

Our device should ideally operate in the regime in which $P$, the probability that at least one photon is emitted, is small. In this case we can write $P \approx 2 \kappa  \tau  \alpha^2 \ll 1$. By substituting the operator $W$ in for $U$ in the expression for $\langle X \rangle$ given in Eq.(\ref{S1}) we can determine the effect of a single emission to first-order in $\gamma$ (because $P$ is already first-order in $\gamma$). Performing this calculation (by permuting the annihilation operator appearing in $W$ through one of the $U$ operators) we find that a single emission, whether it occurs during part I or part III (emissions during part II are already included in the analysis above) effectively shifts the phase of the initial coherent state by a random amount evenly distributed in $[0,\pi]$. The result is that $\alpha$ is symmetrically distributed between the values $\pm\mbox{Re}[\alpha_0]$, thus setting the force signal to zero. This means that for the purposes of measuring the force, $\delta'$, single photon emissions merely reduce the average value of the probability imbalance by a factor of $1-P$, thus inducing a correction to the expression for the signal (Eq.(\ref{xx2})). Given that the peaks of the measured distribution are now at $\pm \xi\eta^2\alpha$, the resulting signal is   
\begin{align} 
  S = \frac{1}{2} - \frac{m}{M} =  2 \alpha' \left[ 1 + \frac{\eta e^{\Gamma T}}{(2\alpha')^2} \right]  \left( 1 - P \right) \delta' ,   \label{Sfull}
\end{align} 
in which $\alpha' \equiv \eta^2 \alpha$ and $P = \pi\kappa\alpha^2/\lambda$. By performing calculations for higher numbers of photon emissions one could, if necessary, obtain the corrections to $S$ to higher order in $P$. A key result of the above analysis is that photon emissions do not have a catastrophic effect on the measurement. What they do is to reduce the quantum advantage, being the linear scaling in $\alpha$.   

We now consider the experimental parameters required to realize the above scheme, while keeping the loss probability, $P$, small. The primary requirement is that the Kerr rate $\lambda$ should be large compared to the photon emission rates $\alpha^2\kappa$ and $\alpha^2\gamma$. In recent experiments~\cite{Teufel11b}, typical values that have been achieved for the mechanical oscillator are $\omega/2\pi = 10~\mbox{MHz}$ with $\gamma/2\pi = 10~\mbox{Hz}$ and an ambient temperature giving $\tilde{n} = 50$. Linear electrical resonators have frequencies of $5-10~\mbox{GHz}$ and damping rates around $\kappa/2\pi =  100~\mbox{kHz}$. As discussed above, in the absence of an oscillator whose nonlinearity can be tuned over a sufficiently large range, we employ a transmission line to couple the Kerr oscillators to the mechanics. The transfer rate, or effective coupling rate, $g$, is then limited by the rate at which the oscillators damp to the transmission lines. A rate of $g/2\pi = 500~\mbox{kHz}$ is certainly achievable~\cite{Teufel11, Lecocq15}. The Kerr nonlinearity can be made quite large ($\sim 50~\mbox{MHz}$) without difficulty, but brings with it an increased loss rate, $\kappa$. Since $\gamma$ is typically mush less that $\kappa$, it the ratio of $\lambda$ to $\kappa$ that is the key limiting factor. To achieve a single photon loss probability equal to $P$ we need this ratio to be $\lambda/\kappa = \pi \alpha^2/P$. Thus to achieve $P \leq 10\%$ for $\alpha = 1.5$ requires $\lambda/\kappa \approx 70$. This would be satisfied, for example, by an oscillator with the pair of parameters $\lambda/2\pi = 7~\mbox{MHz}$ and $\kappa/2\pi = 100~\mbox{kHz}$. With a transfer rate $g/2\pi = 500~\mbox{kHz}$ the loss factors $\gamma \tau$ and $\kappa\tau$ are approximately $10^{-4}$ and $0.2$, respectively, using the loss rates presented above. This collection of parameters would, therefore, allow one to demonstrate an increase of the force signal as a function of $\alpha$, being the signature of the uniquely quantum effect. 


%
%
%


%

\end{document}